\documentclass{article}
\usepackage{amssymb,amsfonts,amsmath,amsthm,lineno}
\usepackage{centernot}
\usepackage[showonlyrefs]{mathtools}
\mathtoolsset{showonlyrefs=true}

\MakeRobust{\eqref}

\usepackage{hyperref}
\usepackage{enumitem}

\newcommand{\mR}{\mathbb{R}}

\newcommand{\kk}{\mathbf{k}}

\newcommand{\lb}{\mathbf{l}}

\newcommand{\yy}{\mathbf{y}}
\newcommand{\vv}{\mathbf{v}}

\newcommand{\rr}{\mathbf{r}}

\newcommand{\la}{\lambda}

\newcommand{\p}{\partial}
\newcommand{\w}{\omega}
\newcommand{\W}{\Omega}

\newcommand{\out}{\mathrm{out}}

\newcommand{\adv}{\mathrm{adv}}

\newcommand{\dV}{\dot{V}}
\newcommand{\ddV}{\ddot{V}}

\newcommand{\ti}{\widetilde}

\DeclareMathOperator{\sgn}{sgn}

\newcommand{\beq}{\begin{equation}}
\newcommand{\eeq}{\end{equation}}

\usepackage[a4paper, total={14.2cm, 25cm}]{geometry}

\title{There is no `velocity kick' memory in electrodynamics}
\author{Andrzej Herdegen\thanks{e-mail: herdegen@th.if.uj.edu.pl}\\
{\it Institute of Physics, Jagiellonian University,}\\
{\it ul.\,S.\,{\L}ojasiewicza 11, 30-348  Krak\'{o}w, Poland}}\date{}

\begin{document}

\maketitle

\begin{abstract}
The memory effect in electrodynamics, as discovered in 1981 by Sta\-rusz\-kie\-wicz, and also analysed later, consists of an adiabatic shift of the position of a test particle. The proposed `velocity kick' memory effect, supposedly discovered recently, is in contradiction to these findings. We show that the `velocity kick' memory is an artefact resulting from an unjustified interchange of limits. This example is a warning against drawing uncritical conclusions for spacetime fields, from their asymptotic behavior.

\end{abstract}

\section{Introduction}

The long range structure of electrodynamics, both classical as quantum, experiences in recent years wide interest of researchers. However, this interest largely ignores, or in some cases is even in contradiction with, earlier knowledge in this field. Some aspects of this problem have been described in \cite{her17}, but here we want to concentrate on the memory effect. In electrodynamics, this effect has been first derived in 1981 by Staruszkiewicz \cite{sta81}, with the use of the semiclassical approximation. Later the effect has been confirmed in several contexts, see \cite{her95}, \cite{her12} and \cite{her17}.

The memory effect described in the above-mentioned works is the result of the interaction of a charged particle with a low energy, free electromagnetic field. The crucial point in this setting is that the field may be chosen (for instance, appropriately scaled) so as to have arbitrarily low energy content, while keeping the long-range spacelike tail (of decay order $r^{-2}$) unchanged. One should expect that such a field cannot change the momentum of the particle, and indeed this has been confirmed. However, it turns out that the interaction is not completely trivial: the trajectory of the particle is adiabatically shifted between far past and far future by a nonzero vector uniquely determined by the long range characteristic of the field. Equivalently, for the wave function of a quantum particle, its phase is shifted by a momentum-dependent quantity.

However, in 2013 appeared an article \cite{bie13}, in which the authors claim to have discovered a `velocity kick', which a test particle will experience in `radiation zone' of an electromagnetic field, both free and with sources. From the measurement of this `kick' one is supposed to be able to uniquely determine the long range characteristics of the field. These new findings are in evident contradiction to the earlier results. At the same time, they found their place in numerous articles gaining wide popularity on the wave of interest mentioned at the beginning (see, e.g., \cite{pas17,psw22}, but in GoogleScholar Ref.\ \cite{bie13} has presently more than 150 other citations). Therefore, it is of primary importance to clarify this matter. We do this here: it is shown that the `kick' interpretation is based on an unallowed mathematical operation, implicitly assumed in this recent analysis.

In order to make this letter self-contained, we gather in Section \ref{free} some facts on solutions of Maxwell's equations and their asymptotic behavior. These properties are discussed in full in \cite{her17}. Section \ref{mem} recalls the shift memory, and then pins down the error in the `kick effect'.

\section{Electromagnetic fields and their asymptotics}\label{free}

The Lorenz potential $A_\mu(x)$ of a field taking part in scattering, and its electromagnetic field $F_{\mu\nu}(x)$, are characterized by having null asymptotes of the following form. There exists vector functions $V_\mu(s,l)$, $V'_\mu(s,l)$, $s\in\mR$, $l\in C_+$, the future light cone, with well defined limit values $V_\mu(\pm\infty,l)$, $V'_\mu(\pm\infty,l)$, related by the condition
\begin{equation}\label{match}
 V_\mu(-\infty,l)=V'_\mu(+\infty,l)\,,
\end{equation}
such that for $l\in C_+$, one has
\begin{gather}
 \lim_{R\to\infty}RA_\mu(x+Rl)=V_\mu(x\cdot l,l)\,,\quad
 \lim_{R\to\infty}RA_\mu(x-Rl)=V'_\mu(x\cdot l,l)\,,\label{Anull}\\
 \lim_{R\to\infty}RF_{\mu\nu}(x+Rl)
 =l_\mu\dV_\nu(x\cdot l,l)-l_\nu\dV_\mu(x\cdot l,l)\,,\label{Fnullfut}\\
 \lim_{R\to\infty}RF_{\mu\nu}(x-Rl)
 =l_\mu\dV'_\nu(x\cdot l,l)-l_\nu\dV'_\mu(x\cdot l,l)\,.\label{Fnullpast}
\end{gather}
where $\dV_\mu(s,l)=\p_sV_\mu(s,l)$. The equality \eqref{match} is the much recently celebrated `matching property', which, however, was obtained much earlier in \cite{her95} (Eq.\ (2.26)). Functions $V_\mu(s,l)$ and $V'_\mu(s,l)$ are homogeneous of degree $-1$ in their arguments (jointly), and satisfy the relations
\begin{equation}
 l\cdot V(s,l)=l\cdot V'(s,l)=q\,,
\end{equation}
where $q$ is the charge of the source of the field. Further more specific restrictions on these functions are discussed in \cite{her17}, but they have no influence on our discussion here.

The (equal) variables of Eq.\ \eqref{match} govern the asymptotic spacelike behavior of the fields: for each spacetime position vector $x$, and each spacelike vector $y$, one has
\begin{gather}
 \lim_{R\to\infty}RA_\mu(x+Ry)
  =\frac{1}{2\pi}\int V_\mu(-\infty,l)\,\delta(y\cdot l)d^2l\,,\label{Asp}\\
  \lim_{R\to\infty}R^2F_{\mu\nu}(x+Ry)
  =\frac{1}{2\pi}
  \int\big[l_\mu V_\nu(-\infty,l)-l_\nu V_\mu(-\infty,l)\big]
  \delta'(y\cdot l)d^2l\,,\label{Fsp}
\end{gather}
where $\delta$ and $\delta'$ are the Dirac delta and its derivative, respectively, and $d^2l$ is the invariant measure on the set of null directions. If one scales null vectors $l$ to $l^0=1$, then $d^2l=d\W(\lb)$, the solid angle measure.

We now restrict attention to the future null infinity, and split the field in the way appropriate for this asymptotic region, into the advanced and the free outgoing parts, $A_\mu(x)=A^\adv_\mu(x)+A^\out_\mu(x)$, and similarly for the field. Let $J_\mu(x)$ be the source current for $A_\mu^\adv(x)$ (with the Maxwell equations in Gauss' units). We denote
\begin{equation}
 V^J_\mu(s,l)=\int \delta(s-x\cdot l)J_\mu(x)dx\,,
\end{equation}
where $\delta$ is the Dirac delta. Then $V_\mu(+\infty,l)=V^J_\mu(+\infty,l)$, and this quantity is related to the potential of the Coulomb field of the outgoing free charged particles.
If we denote further
\begin{equation}\label{Jout}
 V^\out_\mu(s,l)=V_\mu(s,l)-V_\mu(+\infty,l)\,,
\end{equation}
then the future null asymptotes of the advanced, and of the free outgoing fields, are given by
\begin{gather}
 \lim_{R\to\infty}RA^\adv_\mu(x+Rl)=V_\mu(+\infty,l)\,,\quad
 \lim_{R\to\infty}RF^\adv_{\mu\nu}(x+Rl)=0\,,\\
 \lim_{R\to\infty}RA^\out_\mu(x+Rl)=V^\out_\mu(x\cdot l,l)\,,\\
 \lim_{R\to\infty}RF^\out_{\mu\nu}(x+Rl)
 =l_\mu\dV^\out_\nu(x\cdot l,l)-l_\nu\dV^\out_\mu(x\cdot l,l)\,.
\end{gather}
The free outgoing field may be now recovered in the whole spacetime from its null asymptote (this is a special case of the Kirchhoff formula for null initial data):
\begin{gather}\label{out}
 A^\out_\mu(x)=-\frac{1}{2\pi}\int \dV^\out_\mu(x\cdot l,l)\,d^2l\,,\\
 F^\out_{\mu\nu}(x)
 =-\frac{1}{2\pi}
 \int \big[l_\mu\ddV^\out_\nu(x\cdot l,l)
 -l_\nu\ddV^\out_\mu(x\cdot l,l)\big]\,d^2l\,,\label{outF}
\end{gather}
where $\ddV_\mu(s,l)=\p_s^2V_\mu(s,l)$. The above integral representation is related to the more standard Fourier representation
\begin{equation}\label{Aint}
 A^\out_\mu(x)=\frac{1}{\pi}\int e^{-ix\cdot k}a_\mu(k)\sgn(k^0)\delta(k^2)d^4k
\end{equation}
by
\begin{equation}\label{aV}
 \w a_\mu(\w l)=-\ti{\dV^\out_\mu}(\w,l)
 =-\frac{1}{2\pi}\int e^{i\w s}\dV^\out_\mu(s,l)ds\,.
\end{equation}

\section{Electromagnetic memory effects}\label{mem}

In 1974 Zel'dovitch and Polnarev have shown \cite{zel74} that a pair of test bodies exposed to a gravitational wave burst experiences a finite and permanent relative position shift between remote past and future. This effect was discussed again and named the `memory' of the gravitational wave-burst in 1985 \cite{bra85}.

Strangely enough, only after the work of 1974 in gravitation, was an analogous memory effect discussed in electrodynamics, where the problem may be formulated in very clear terms referring to a laboratory experiment. Staruszkiewicz \cite{sta81} was the first author to pose in 1981 a question: does a (free)  electromagnetic field in zero frequency limit produce observable effects? What is meant by zero frequency is the following. Let $A_\mu(x)$ be a Lorenz potential of a field from the class identified in the last section, and define its rescaled version $A_\mu^{(\la)}(x)=\la^{-1}A_\mu(\la^{-1}x)$. Consider the scaled field in the large $\la$ limit. It is easy to see that the energy carried by this field vanishes in that limit, so it is unable to change the velocity of any massive charged particle. In terms of the Fourier transform, $\ti{\dV_\mu^{(\la)}}(\w,l)=\ti{\dV_\mu}(\la\w,l)$, so the frequency content shrinks to values around $\w=0$, but the spacelike tail due to $V_\mu(-\infty,l)=-2\pi\ti{\dV_\mu}(0,l)$ (see \eqref{Asp} and \eqref{aV}) does not change in the limit. Using the semiclassical approximation
for the phase of the wave function of a test particle placed in such field,  Staruszkiewicz found that the incoming plane wave $\exp(-ip\cdot x)$ of a charged particle, with $e$ and $p$ its charge and four-momentum, respectively, acquires in far future a phase shift of the magnitude
\begin{equation}\label{phase}
 \delta(p)=-\frac{e}{2\pi}\int\frac{p\cdot V(-\infty,l)}{p\cdot l}\,d^2l\,.
\end{equation}
If a wave packet is formed, this shift produces observable effects. It is easy to see their nature. If $f(p)$ is the momentum profile of the initial packet, then the final packet has the profile $e^{i\delta(p)}f(p)$. The addition to the phase has no effect on the distribution of momentum, but under the action of the position operator $-i\p/\p p^\mu$ causes a shift $\p\delta(p)/\p p^\mu$.

This effect has been later confirmed in other contexts. The same trajectory shift has been derived for a classical test particle in \cite{her95}, later reproduced with another method in \cite{her17}. The memory effect for the Dirac field, both classical as quantum, placed in low-energy electromagnetic field was analyzed in \cite{her12}. For the quantum Dirac field one finds the scattering operator in the form
\begin{equation}
 S=\exp\Big[i\int\delta(p)\rho(p)d\mu(p)\Big]\,,
\end{equation}
where $\delta(p)$ is the phase \eqref{phase}, $\rho(p)$ is the momentum-density of the charge operator and $d\mu(p)$ is the invariant measure on the mass hyperboloid.

All instances of the memory effect described above have two characteristic properties:\\
(i) energy involved in causing them is zero (in the limit), and\\
(ii) they are exclusively due to the long-range characteristic of the electromagnetic free field $V(-\infty,l)$, on which spacelike tails depend, Eqs.\ \eqref{Asp} and \eqref{Fsp}.

In 2013 a new article on electromagnetic memory effect \cite{bie13} appeared, whose conclusions gained wide popularity. The authors claim to have related  asymptotic characteristics of the electromagnetic fields, both free as well as produced by sources, to a `velocity kick', which supposedly a test particle will experience if it is placed in the `radiation zone'. To obtain the magnitude of this `kick', it is postulated that the electromagnetic field should be integrated over time (more precisely, the authors refer to `electric field', but the extension doesn't change anything in our discussion). For this purpose the spacetime position vectors $x$ are parametrized by the retarded coordinates, that is $x=ut+Rk$, where $t$ is the unit time axis vector, $k$ is a future null vector such that $t\cdot k=1$, $R\geq0$, and then $u$ is the retarded time. It is postulated that for a given $k$ the observation is made at a point with large $R$, where the field is weak. For large $R$, the authors further argue---and this is the crucial point---one can take the leading order of the field, which in our language is given by the limit \eqref{Fnullfut}. Thus, apart from the factor $R^{-1}$, the integral of the field over time $u$ is given in the leading order by
\begin{equation}\label{int}
\begin{aligned}
 \int_\mR\lim_{R\to\infty}RF_{\mu\nu}(ut+Rk)\,du
 &=\int_\mR\big[k_\mu\dV_\nu(u,k)-k_\nu\dV_\mu(u,k)\big]du\\
 &=k_\nu V^\out_\mu(-\infty,k)-k_\mu V^\out_\nu(-\infty,k)\,,
\end{aligned}
\end{equation}
where we used \eqref{Fnullfut}, and then \eqref{Jout}. The rhs depends on the long-range variable of the outgoing field, that is the difference $V_\mu(-\infty,k)-V_\mu(+\infty,k)$, which appears in recent discussions of memory. However, this calculation does not do justice to the actual experiment proposition. However large $R$ is, it is for a concrete measurement  experiment fixed. And then, as it turns out, the above leading order calculation gives a totally false result. Consider the free outgoing field first. For any $R$, the field $F^\out_{\mu\nu}(ut+Rk)$ is given by \eqref{outF}, and a simple calculation gives
\begin{equation} \label{Int}
\int_\mR F^\out_{\mu\nu}(ut+Rk)du=0\,,
\end{equation}
as $\dV_\mu(\pm\infty,k)=0$. The `kick' effect does not exist for this, and any free field, not only in the radiation zone, but anywhere. On the other hand, the advanced field does not contribute to the formula \eqref{int} at all, but using the advanced Green function one finds
\begin{equation}\label{Intadv}
 \int_\mR RF^\adv_{\mu\nu}(ut+Rk)du
 =\int\frac{R}{|\yy-R\kk|}\big[\p_\mu J_\nu(y)-\p_\nu J_\mu(y)\big]dy\,.
\end{equation}
This is nonzero in general, and may even have a finite, nonzero limit for $R\to\infty$. However, this limit, even if it exists, has in general nothing to do with the memory variables. For example, an easy calculation shows that for the advanced field $F_{\mu\nu}^C(x)$ (i.e.\ the Coulomb field, in this case) of a particle with charge $q$,  moving freely along the trajectory $z^\mu(\tau)=z_0^\mu+\tau v^\mu$, one has
\begin{equation}
 \int_\mR RF^C_{0i}(ut+Rk)du
 =2q\frac{v_0 r_\bot^i}{|\vv||\rr_\bot|^2}\,,
\end{equation}
where $\rr_\bot$ is the orthogonal to $\vv$ part of the vector $\rr=\kk-R^{-1}\mathbf{z}_0$.
This not only grossly disagrees with \eqref{int}, but it also falsifies the `slow motion case' analysis of \cite{bie13}, in which the authors keep, as we have done here, fixed $R$. The reason for this discrepancy is that the authors apply the dipole approximation \cite{lan80}, which does not work here. It works well for sources restricted to bounded regions in space and finite times, details may be found in the textbook. For infinite time integration even a slow free asymptotic motion of charges involves infinite displacements. Let us also note, that the result of the slow motion analysis in \cite{bie13} involves only the in and out velocities of charges, so it disagrees with \eqref{int} as well.

What is the mechanism of the discrepancy between \eqref{int} on the one hand, and  \eqref{Int} and \eqref{Intadv} on the other? It is the unallowed pulling of the limit sign under the integral in \eqref{int}. One can show that $RF^\out_{\mu\nu}(ut+Rk)$ cannot be bounded, uniformly in $R$, by an integrable function of $u$, so the usual tool of the dominated convergence theorem does not apply here. The mechanism is similar as for an integral $\int_\mR [f(u)-f(u+R)]du$, where $f$ may be, for simplicity, any continuous function of compact support. The integral vanishes, but the integral of the point-wise limit of the integrand is $\int_\mR f(u)du$. The precise mechanism for the advanced field is slightly different, but the conclusion on the discrepancy is the same.
We conclude that the observation of the velocity kick misses the original motivation. We also note that this example illustrates a general warning against drawing unjustified conclusions for the `bulk' properties of fields, from their asymptotic behavior.

However, this does not prohibit independent determination of $V_\mu(\pm\infty,l)$. As we have seen, $V_\mu(-\infty,l)$ is encoded in the spacelike asymptotic behavior of fields \eqref{Asp}, \eqref{Fsp}, while $V_\mu(+\infty,l)$ in the asymptotic trajectories of the outgoing charges.
The important aspect of the adiabatic memory shift discovered earlier is, that for a weak free field it allows to observe its spacelike tail in scattering experiments, without going to spatial infinity (if the amplitude is measured, and not merely the cross section).

To better understand the clash between the adiabatic shift memory and the identity  \eqref{Int}, we reproduce the shift of the trajectory of a charged particle in a weak free field, say $F^\out_{\mu\nu}$,~\cite{her17}. For a particle passing through a spacetime point $x_0$, with the four-velocity $v$, which is not affected in the low energy of the field limit, the shift is given by
\begin{align}
 \Delta_\mu&=-\frac{e}{m}\int_\mR F^\out_{\mu\nu}(x_0+v\tau)\tau d\tau\,v^\nu\\
 &=\frac{e}{2\pi m} \int
 \frac{l_\mu V^\out_\nu(-\infty,l)-l_\nu V^\out_\mu(-\infty,l)}{(v\cdot l)^2}
 d^2l\,v^\nu\,,
\end{align}
where the first equality follows from an analysis of the equation of motion, and for the second equality we used \eqref{outF}. The result agrees with the derivative of~$\delta(p)$.

Finally, we comment on the idea of effectively restricting the integration domain in the retarded time to a compact interval, an issue acknowledged in part of the literature on the `kick memory', see e.g.\ \cite{pas17}. First of all, integrating $F_{\mu\nu}(ut+Rk)$ for any finite $R$ over a compact interval $[u_1,u_2]$ one obtains an accidental quantity, not related to asymptotic degrees of freedom. Moreover, for any integration interval this quantity vanishes in the limit $R\to\infty$, so there is no `kick'. In spite of that, let us multiply this quantity by $R$, and take the limit. The limit commutes with compact integration, so one finds
\begin{multline}
 \lim_{R\to\infty}\int_{u_1}^{u_2}RF_{\mu\nu}(ut+Rk)du
 =\int_{u_1}^{u_2}\lim_{R\to\infty}RF_{\mu\nu}(ut+Rk)du\\
 =k_\nu[V_\mu(u_1,k)-V_\mu(u_2,k)]-k_\mu[V_\nu(u_1,k)-V_\nu(u_2,k)]\,.
\end{multline}
However, with $\Delta V(k)=V(-\infty,k)-V(+\infty,k)$ fixed, the function $V(u,k)$ may be arbitrarily slowly varying, and the rhs is then arbitrarily small. The point of an experimental memory effect is to produce $\Delta V(k)$, whatever the rate of change. Thus, not knowing the field \emph{a~priori}, one has to take in the above identity the limits $u_1\to-\infty$ and $u_2\to+\infty$, which brings us back to the relation \eqref{int}, critically considered earlier.

Let us stress once more that the proper setting for the analysis of electromagnetic effects is that of laboratory experiments, so for any $R$ extended integration in time leads out of `radiation zone'. In this respect the situation is different than in gravitation, where cosmic scales are involved, and in many instances one can argue that all history of humankind is contained in `radiation zone' of some radiation events.


\frenchspacing

\begin{thebibliography}{00}
\bibitem{her17} Herdegen, A.: Asymptotic structure of electrodynamics
    revisited, Lett. Math. Phys. {\bf 107}, 1439-1470 (2017).
    https://doi.org/10.1007/s11005-017-0948-9
\bibitem{sta81} Staruszkiewicz, A.: Gauge invariant surface contribution to the number of photons integral, Acta. Phys. Pol. B {\bf 12}, 327-337 (1981)
\bibitem{her95} Herdegen, A.: Long-range effects in asymptotic fields
    and angular momentum of classical field electrodynamics, J. Math. Phys.
   {\bf 36}, 4044-4086 (1995)
\bibitem{her12} Herdegen, A.: Infrared limit in external field scattering, J. Math. Phys. {\bf 53},  052306 (2012)
\bibitem{bie13} Bieri, L. and Garfinkle, D.: An electromagnetic analogue of gravitational wave memory, Class. Quant. Grav. {\bf 30}, 195009 (2013)
\bibitem{pas17} Pasterski, S.: Asymptotic symmetries and electromagnetic memory, JHEP {\bf 09}, 154 (2017)
\bibitem{psw22} Prabhu, K., Satishchandran, G. and Wald, R.M.: Infrared finite scattering theory in quantum field theory and quantum gravity, Phys. Rev. D {\bf 106}, 066005 (2022)
\bibitem{zel74} Zel'dovitch, Ya.B and Polnarev, A.G.: Radiation of gravitational waves by a cluster of superdense stars. Astron. Zh. {\bf 51}, 30-40 (1974) [So. Astron. {\bf 18}, 17-23 (1974)]
\bibitem{bra85} Braginsky, V.B. and Grishchuk, L.P.: Kinematic resonance and memory effect in free mass gravitational antennas, Zh. Eksp. Teor. Fiz. {\bf 89}, 744-750 (1985) [Sov. Phys. JETP {\bf 62}, 427-430 (1985)]
\bibitem{lan80} Landau, L.D. and Lifshitz, E.M.: {\it The Classical Theory of Fields}, 4\textsuperscript{th} ed., Butterworth-Heinemann, Amsterdam, 1980





\end{thebibliography}
\end{document}